\documentstyle[amsfonts,amsbsy,epsf,12pt]{article}
\def\titlematter{
\begin{center}
  {\bf NON-EQUILIBRIUM DESCRIPTION\\
       OF BREMSSTRAHLUNG IN DENSE MATTER\\
       (Landau - Pomeranchuk - Migdal Effect)}
  \footnote{\em GSI-preprint 95-18; submitted to Phys. Lett. B\\[-2mm]}\\[5mm]
  {J\"orn Knoll and Dmitri N. Voskresensky
  \footnote{permanent adress: Moscow Institute for Physics and
  Engineering, Russia, 115409 Moscow, Kashirskoe shosse 31}}\\[3mm]
  {\it Gesellschaft f\"ur Schwerionenforschung (GSI)\\
       Postfach 110 552 \\
       D-64220 Darmstadt, Germany}
\end{center}
              }
\def\eqs{\makebox(0,0){$=$}}
\def\pls{\makebox(0,0){$+$}}

\def\ssp{\makebox(0,0)
    {\thinlines\put(-.1,0){\line(1,0){.2}}\put(0,-.1){\line(0,0){.2}}}}
\def\ssm{\makebox(0,0){\put(-.1,0){\thinlines\line(1,0){.2}}}}
\def\photon{\thinlines\multiput(0,0)(.2,0){3}{\line(1,0){0.1}}}

\def\vBoson{\thicklines
           \multiput(0.,-.3125)(.0,.25){4}{\oval(.125,.125)[r]}
           \multiput(0,-.4375)(.0,.25){4}{\oval(.125,.125)[l]}
           }
\def\vhBoson{\thicklines
           \multiput(-.5,-.375)(.25,.25){4}{\oval(.25,.25)[rb]}
           \multiput(-.25,-.375)(.25,.25){4}{\oval(.25,.25)[lt]}
           }
\def\hvBoson{\thicklines
           \multiput(.5,-.375)(-.25,.25){4}{\oval(.25,.25)[lb]}
           \multiput(.25,-.375)(-.25,.25){4}{\oval(.25,.25)[rt]}
           }

\def\Fermion{\thicklines\put(0,0){\vector(1,0){.6}}
            \put(0.6,0){\line(1,0){.4}}}
\def\fermion{\thinlines\put(0,0){\vector(1,0){.6}}
            \put(0.6,0){\line(1,0){.4}}}

\def\oneloop{
     \put(1.5,0){\thicklines\oval(2.0,1.5)}
     \put(0,0){\photon}\put(0.3,0.3){\ssp}
     \put(2.5,0){\photon}\put(2.7,0.3){\ssm}}
\def\fullself{\begin{picture}(5,1)\put(0,0){\oneloop}
     \put(1.5,0){\makebox(0,0){$-\ii{\boldsymbol \Pi}$} }
     \end{picture}}
\def\oneloopvertex{
    \put(1.625,0){\thicklines\oval(2.0,1.5)}
    \put(0,0){\photon}\put(0.35,0.3){\ssp}
    \put(0.625,0){\circle*{.25}}\put(2.625,0){\circle*{.25}}
    \put(2.75,0){\photon}\put(2.9,0.3){\ssm}}

\def\mediumloop{
    \put(2.125,0){\thicklines\oval(3.0,1.5)}
    \put(0,0){\photon}\put(0.35,0.3){\ssp}
    \put(0.625,0){\circle*{.25}}\put(3.625,0){\circle*{.25}}
    \put(3.75,0){\photon}\put(3.9,0.3){\ssm}}

\def\doubleloop{
    \put(1.125,0){\thicklines\oval(1,2)}\put(2.625,0){\thicklines\oval(1,2)}
    \put(0,0){\photon}\put(0.35,0.3){\ssp}
    \put(0.625,0){\circle*{.25}}\put(3.125,0){\circle*{.25}}
    \put(3.25,0){\photon}\put(3.4,0.3){\ssm}
    \multiput(1.875,-.4)(0,.8){2}{\makebox(0,0){\rule{3mm}{1.5mm}}}
    \put(1.875,-.8){\ssm}\put(1.875,.8){\ssp} }
\def\Doubleloop{
    \put(1.125,0){\thicklines\oval(1,2)}\put(3.625,0){\thicklines\oval(1,2)}
    \put(0,0){\photon}\put(0.35,0.3){\ssp}
    \put(0.625,0){\circle*{.25}}\put(4.125,0){\circle*{.25}}
    \put(4.25,0){\photon}\put(4.4,0.3){\ssm}
    \multiput(1.875,-.4)(0,.8){2}{\makebox(0,0){\rule{3mm}{1.5mm}}}
    \multiput(2.875,-.4)(0,.8){2}{\makebox(0,0){\rule{3mm}{1.5mm}}}
    \multiput(2.375,-.4)(0,.8){2}{\thicklines\oval(.5,.7)}
    \put(1.875,-.8){\ssp}\put(1.875,.8){\ssp}
    \put(2.875,-.8){\ssm}\put(2.875,.8){\ssm} }

\def\Fullbox{\thicklines\put(0,0){\vector(1,0){.5}}\put(.5,0){\vector(1,0){1}}
                \put(0,1){\vector(1,0){.5}}\put(.5,1){\vector(1,0){1}}
                \put(.75,.5){\makebox(0,0){\rule{.5cm}{1.cm}}}
                \put(.75,1.2){\ssm}\put(.75,-.2){\ssm}}
\long\def\fourpointeq{\begin{picture}(14,1.2)
     \put(0,.25){\Fullbox}\put(2,.75){\eqs}
     \put(2.5,.25){\thicklines\put(0,0){\vector(1,0){.625}}
              \put(0,1){\vector(1,0){.625}}
              \put(.875,0){\vector(1,0){.625}}\put(.875,1){\vector(1,0){.625}}
              \put(.75,.5){\vBoson}
              \put(.75,0){\circle*{.2}}\put(.75,1){\circle*{.2}}
              \put(.75,1.2){\ssm}\put(.75,-.2){\ssm}}
     \put(4.5,.75){\pls}
     \put(5,.25){\thicklines\put(0,0){\vector(1,0){.625}}
           \put(0,1){\vector(1,0){.625}}
           \put(.875,0){\vector(1,0){1.625}}\put(.875,1){\vector(1,0){1.625}}
           \put(.75,1.2){\ssm}\put(.75,-.2){\ssm}
           \put(1.75,1.2){\ssm}\put(1.75,-.2){\ssm}
           \put(.75,.5){\vBoson}\put(1.75,.5){\vBoson}
           \put(.75,0){\circle*{.2}}\put(.75,1){\circle*{.2}}
           \put(1.75,0){\circle*{.2}}\put(1.75,1){\circle*{.2}}  }
     \put(8,.75){\pls}
     \put(8.5,.25){\thicklines\put(0,0){\vector(1,0){.625}}
           \put(0,1){\vector(1,0){.625}}
           \put(.875,0){\vector(1,0){1.625}}\put(.875,1){\vector(1,0){1.625}}
           \put(1.25,.5){\vhBoson}\put(1.25,.5){\hvBoson}
           \put(.75,0){\circle*{.2}}\put(.75,1){\circle*{.2}}
           \put(1.75,0){\circle*{.2}}\put(1.75,1){\circle*{.2}}
           \put(.75,1.2){\ssm}\put(.75,-.2){\ssm}
           \put(1.75,1.2){\ssm}\put(1.75,-.2){\ssm}}
     \put(12,.75){\makebox(0,0){$\dots$}}\end{picture}}

\def\Dysonf{\begin{picture}(8,1)
\put(0,0){\Fermion}\put(1.5,0){\eqs}\put(2,0){\fermion}\put(3.5,0){\pls}
\put(4,0){\fermion}\put(5.75,0){\thicklines\oval(1.5,1.0)}
\put(5.75,0){\makebox(0,0){$-i{\boldsymbol \Sigma}$}}\put(6.5,0){\Fermion}
        \end{picture} }

\def\fullbox{\makebox(0,0){\rule{1.5mm}{3mm}}}
\def\interaction{\makebox(0,0){\put(0,0){\interact}
    \put(0,.95){\ssp}\put(0,-.95){\ssm}
    \put(0,.125){\ssp}\put(0,-.125){\ssm}}}
\def\interact{\makebox(0,0){\put(0,.5){\fullbox}
    \thicklines\put(0,0){\oval(.75,.5)}
    \put(0,-.5){\fullbox}} }

\def\til2loop{\put(0,0){\oneloopvertex}\put(4.,0){\pls}
      \put(4.75,0){\oneloopvertex}\put(6.375,0){\interaction}
      \put(9,0){\pls}}

\def\keydiagrams{\begin{picture}(21,4)\put(0,3){
   \put(0,0){\fullself}\put(3.5,0){\eqs}\put(4,0){\til2loop}}
   \put(13.75,3){\put(0,0){\mediumloop}
      \multiput(1.625,0)(1,0){2}{\interaction}}
      \put(18.75,3){\pls}\put(19.75,3){\makebox(0,0){$\dots$}}
   \put(-1,0){
   \put(4,.2){\doubleloop}\put(8.5,.2){\pls}\put(3.25,.2){\pls}
   \put(9.25,.2){\put(0,0){\mediumloop}
      \thicklines\multiput(1.615,-.75)(.02,0){10}{\line(2,3){1}}
      \multiput(2.615,-.75)(.02,0){10}{\line(-2,3){1}}
      \put(1.625,.95){\ssp}\put(1.625,-.95){\ssm}
      \put(2.625,.95){\ssm}\put(2.625,-.95){\ssp}}
   \put(14.25,.2){\pls}\put(15,.2){\Doubleloop}\put(20.5,.2){\pls}
   \put(21.5,.2){\makebox(0,0){$\dots$}}}
   \end{picture}}

\newcommand{\yourabstract}[1]{\begin{footnotesize} \mbox{}\\
 {\bf\noindent Abstract:} \begin{center} \mbox{}\parbox[t]{5.in}{#1}
 \end{center}\end{footnotesize} }
\newcommand{\yoursection}[1]{ $ $\\ {\bf\noindent #1}\\[2mm]}

\newcommand{\di}{{\mathrm d}}
\newcommand{\ii}{{\mathrm i}}

\renewcommand{\Re}{{\bf Re}}
\renewcommand{\Im}{{\bf Im}}
\begin{document}$ $\\
\titlematter
\yourabstract
{The soft behavior of the bremsstrahlung from a source is discussed in
terms of classical transport models and within a non--equilibrium
quantum field theory (Schwinger - Kadanoff - Baym -
Keldysh) formulation.}

\yoursection{Introduction}
We study the importance of coherence time effects (Landau -
Pomeranchuk - Migdal - effect (LPM)\footnote{The original LPM
considerations were restricted to cases where ultra-relativistic
particles traverse a finite piece of target matter. }\cite{LPM}) on
the production and absorption of field quanta from the motion of
source particles in non-equilibrium dense matter. In order to
calculate such bremsstrahlung effects appropriately one needs to go
beyond the commonly used quasi--particle picture, and include effects
of the finite widths of the particles \cite{K,KL,WG}.  Our
considerations are of particular interest for the application to
photon, or di-lepton production in high energy nuclear collisions, for
gluon or parton radiation and absorption in QCD transport and its
practical implementation in parton kinetic models (such problems are
discussed, e.g. in \cite{WG, Proc}), to neutrino and axion radiation
from supernovas and neutron-star matter (see \cite{NS}), for the soft
phenomena in quantum cosmological gravity (see \cite{TW}) and also for
many condensed matter phenomena, as particle transport in metals and
semiconductors, radiation in plasma etc (see \cite{SL}).  The study
also gives some hints how to generalize the standard transport picture
(see \cite{D}) such that it can include genuine off-shell effects due
to damping of the single particle propagation from collisions or
decays.

To be specific we take the example of electrodynamics, considering
photon production and assume that the source system couples only
perturbatively (to lowest order) to the radiation field, while the
source itself can interact in any non-perturbative way. Thus, we
consider a "white" body as a source for the radiated field!  Our
considerations are formulated in real time non-equilibrium Green's
function technique, where the production rate is given by the $-+$
component of the proper self-energy diagram of the produced photon
\begin{eqnarray} \label{Pi-+xq}\unitlength6mm
   -\ii\Pi^{-+}= \begin{picture}(4.3,.7) \put(0.2,0.2){\fullself}
   \put(3.8,0.2){\eqs} \end{picture} 4\pi\int d^4 \xi e^{\ii
   q\xi}\left<j^{\nu\dagger}(x-\xi/2) j^\mu(x+\xi/2)\right>
   ,
\end{eqnarray}
which is determined by the current auto correlation function of the
source. The dashed lines relate to the photon, while the
$-\ii\Pi$-loop symbolically denotes the exact inclusion of all strong
interaction among the source particles. The bracket
$\langle\dots\rangle$ denotes a quantum ensemble average over the
source with quantum states and operators in the interaction picture.
Above and throughout later we use the Keldysh $\{-,+\}$ notation, see
\cite{LP}, where the $\Pi^{-+}$ and $ \Pi^{+-}$ self energies are
responsible for gain and loss.  Such a formalism has been applied in
calculation of neutrino emissivity of neutron stars in ref.\cite{V1},
employing the quasi--particle approximation for the equilibrium
nucleon Green functions. However the general formalism allows to go
beyond this limit and to account for the finite damping width of the
source particles due to their finite mean free path.

\yoursection{Classical radiation}
For some classical source systems perturbatively coupled to a boson
field the boson self energy can be obtained in closed form. For these
cases one has to evaluate the current-current correlator on the
classical level. The properties of classical radiation will be
illustrated for two examples, where a charged particle (the source)
stochastically moves in dense matter. Thereby the motion of the charge is
described either ($a$) by mesoscopic transport (diffusion process) or
($b$) by a microscopic Langevin process.\\

\noindent{\em Diffusion process}

The motion of the source particle is assumed to be described by a time
dependent phase-space distribution $f({\vec x},{\vec v},t)$ in space
and velocity with convective current density ${\vec j}({\vec x},t)= e
\int \di^3 v\;{\vec v}\;f({\vec x},{\vec v},t)$.  For standard
dissipative media in equilibrium the velocity autocorrelation function
(integrated over space) decays exponentially in time\footnote{We
should note that this ansatz ignores finite size corrections in terms
of some anti-correlations on the scale of a mean recurrence time \cite{KL}.}

\begin{equation}\label{vrelax} \left< v^i(\tau)
    v^k(0)\right>=\frac {1}{3} \left< {\vec v}^2\right> \delta^{ik}
    e^{-\Gamma_x |\tau|},
\end{equation}
where $i$ and $j$ denote the spatial components. $\Gamma_x$ is the
relaxation rate which we approximate as constant,
$\langle\dots\rangle$ now denotes the average over the classical
distribution functions.

Both $f({\vec x},{\vec v},t)$ and the autocorretion function can be
obtained in closed form, if the time evolution of $f$, and the
propagation of fluctuations $\delta f$ are governed by a standard
diffusion process.  Solving then a {\em Fokker-Planck} equation for
the space-time dependence of velocity fluctuations we obtain (in mixed
$\tau,{\vec q}$ representation)
\begin{eqnarray}\label{Picltau}
-\ii\Pi_{\mathrm{cl}}^{-+}(\tau, {\vec
    q})&=&4\pi\int \di^3 {\vec x} e^{-\ii{\vec q}{\vec x}}
    \left<j^i({\vec
    x},\tau)j^k({\vec 0},0)\right> \nonumber \\
    &=&4\pi e^2\rho_0\left\{\left<\left<  v^iv^k \right>\right>_{eq}
    e^{-\Gamma_x|\tau|} -D^2 q^iq^k\left(e^{-\Gamma_x|\tau|}-1\right)^2
    \right\} \\
&&\times\exp\left\{-\frac{D\,{\vec q}^2}{\Gamma_x}
    \left(\Gamma_x|\tau|+e^{-\Gamma_x|\tau|}-1\right)\right\}.
\nonumber
\end{eqnarray}
Here the ensemble average
\mbox{$\langle\dots\rangle_{eq}$}
over the equilibrium distribution $f_{eq}$ keeps only even moments of
${\vec v}$ with \mbox{$\langle{\vec v}^2\rangle= 3D\Gamma_x$} such
that only even powers of $i{\vec q}$ survive.  This renders
$-\ii\Pi_{\mathrm{cl}}^{-+}(\tau,{\vec q})$ real and symmetric in
$\tau$.  For transverse photons ${\vec
\eta}{\vec q}=0$ some terms drop and we have
\begin{eqnarray}\label{Mclres}
    &-\ii\Pi_{\mathrm{cl}}^{-+}(\omega,{\vec q}\,)
    &=4\pi e^2 \rho_0\left< v^iv^j\right>_{eq}
     \exp\left[{D\,{\vec q}^2}/{\Gamma_x}\right]\cr &&\times
   \sum_{k=0}^\infty \frac{1}{k!}
   \left(\frac{-D\,{\vec q}^2}{\Gamma_x}\right)^k
     \frac{2(k+1)\Gamma_x+2D\,{\vec q}^2}
     { \left((k+1)\Gamma_x+D\,{\vec q}^2\right)^2+\omega^2}\\
   &&\longrightarrow 4\pi e^2 \rho_0\left<
   v^iv^j\right>_{eq}2\Gamma_x/(\omega^2+\Gamma_x^2){\quad}
   {\mathrm{for}}\quad \langle v^2\rangle{\vec q}^2\ll\Gamma^2_x.
\end{eqnarray}
Compared to the infra--red divergent quasi--free result $\propto
1/\omega^2$ this form of the correlation renders the photon self
energy regular in the soft limit at four momentum $q=0$. It is
determined by mesoscopic transport properties, namely by the
relaxation rate $\Gamma_x$ and the diffusion coefficient $D=\left<
{\vec v}^2 \right>/3\Gamma_x$; $\rho_0$ is the spatial density of the
charged particles.  For large $q$, i.e. $Dq^2\gg \Gamma_x$, only the
short time behavior of the autocorrelation matters and from
(\ref{Picltau}) and (\ref{Mclres}) one finds in ($\tau , \vec{q}$) and
($\omega , \vec{q}$) representations
\begin{eqnarray}\label{Dq}
\lim_{D\,{\vec q}^2\gg\Gamma_x} \left[-\ii\Pi_{\mathrm
    cl}^{-+}(\tau,{\vec q}\,)\right]
    &=&4\pi e^2\rho_0\left< v^iv^k\right>_{eq}
    \exp\left[-\Gamma_x|\tau|-D\,{\vec q}^2\Gamma_x\tau^2/2\right]
  \, , \\\label{Dqomega}
    \lim_{D\,{\vec q}^2\gg\Gamma_x} \left[-\ii\Pi_{\mathrm
    cl}^{-+}(\omega,{\vec q}\,)\right]
    &=&4\pi e^2\rho_0\left< v^iv^k\right>_{eq}
    \left(\frac{2\pi}{D\,{\vec q}^2\Gamma_x}\right)^{3/2}
    \exp\left\{-\frac{\omega^2}{2D\,{\vec q}^2\Gamma_x}\right\}\,.
    \end{eqnarray}

\noindent{\em Microscopic Langevin process}

In a {\em microscopic Langevin} process hard scatterings occur at
random with a constant mean collision rate $\Gamma$. These scatterings
consecutively change the velocity of a point charge from ${\vec v}_m$
to ${\vec v}_{m+1}$,$\dots$ (subscripts $m$, and $n$ below refer to
the collision sequence). In between scatterings the point charge moves
freely. For such a multiple collision process one can at least
determine the ${\vec q}=0$ part of the self energy.

The modulus of the autocorrelation function takes a Poissonian form
for such a collision sequence
\begin{equation}\label{Apoisson}
   -\ii\Pi_{\mathrm{cl}}^{-+}(\tau,{\vec q}=0)=4\pi e^2\rho_0
   \left<v^i(\tau)v^k(0)\right>
=4\pi e^2\rho_0e^{-|\Gamma\tau|}\sum_{n=0}^\infty
    \frac{|\Gamma\tau|^n}{n!} \left< v^i_m v^k_{m+n}\right>_m,
    \end{equation}
where $\langle\dots\rangle_m$ denotes the average over the discrete collision
sequence $\{m\}$. The time Wigner transform of (\ref{Apoisson}) determines
the spectrum at vanishing $\vec q$ for all $\omega$
\begin{eqnarray} \label{Mtau}
    -\ii\Pi_{\mathrm{cl}}^{-+}(\omega,{\vec q}=0)
    =4\pi e^2\rho_0 \sum_{n=0}^\infty
     \left< v_m^i v_{m+n}^j\right>_m
    \frac{2\Gamma^n{\mathrm{Re}} \left\{ (\Gamma+\ii\omega )^{n+1}\right\} }
    {(\omega^2+\Gamma^2)^{n+1}} .
\end{eqnarray}
This is a genuine {\em multiple collision} description for the photon
production rate in completely regular terms due to the
\mbox{$(\omega^2+\Gamma^2)^n$} form of all denominators. Each term is
regular, since right from the beginning one accounts for the damping
of the source particle because of the finite mean time $1/\Gamma$
between collisions. The result (\ref{Mtau}) still accounts for the
{\em coherence} of the photon field, now expressed through the
correlations $\left<{\vec v}_m{\vec v}_{m+n}\right>_m$ arizing from
the sequence of collisions. The terms in (\ref{Mtau}) define partial
rates, which are associated with specific self energy diagrams.

For the Langevin process the ${\vec q}$-dependence of the self energy
cannot be obtained in closed form in general. Still the $n=0$ term
(c.f. with $n=0$ term from eq.(\ref{Mtau})) can be given
\begin{equation}\label{Mtaun0}
-\ii\Pi_{\mathrm{cl}}^{-+}(\omega,{\vec q})
    \approx 4\pi e^2\rho_0
     \left< \frac{2\Gamma \,v_m^i v_m^k}{(\omega -{\vec q}{\vec v})^2
    +\Gamma^2}\right>_m\,.
\end{equation}
It shows the typical Cherenkov enhancement at $\omega={\vec q}{\vec
v}$. Although for \mbox{${\vec q}\rightarrow 0$} the analytical form of
(\ref{Mtaun0}) resembles the diffusion result (\ref{Mclres}), it is
not the same unless
$\left<{\vec v}_m\;{\vec v}_{m+n}\right>_m=0$ for $n\ne 0$,
an approximation recently used in ref. \cite{Cleymans}. In the general
case velocity correlations between successive scatterings exist, and
there will be a sizeable difference between the microscopic mean
collision rate $\Gamma$ and the mesoscopic relaxation rate
$\Gamma_x$. For systems, where the velocity is degraded by a constant
fraction $\alpha$, such that $\left< {\vec v}_m\cdot{\vec v}_{m+n}
\right>_m= \alpha^n \left< {\vec v}_m\cdot{\vec v}_{m} \right>_m$,
one can sum up the whole series in (\ref{Mtau}) and recover the
diffusion result (\ref{Mclres}) at ${\vec q}=0$ with
$\Gamma_x=(1-\alpha)\Gamma$. This clarifies that the diffusion result
(\ref{Mclres}) represents a resummation of the random multiple
collision result (\ref{Mtau}).\\

\noindent{\em Monte Carlo evaluation of amplitudes}

Some of the cascade schemes try to cure the infra-red problem by
considering the phase of the photon field along the classical orbits.
Thus they evaluate $\int \di t {\vec v}(t)\exp[\ii\omega t-\ii{\vec
q}{\vec x}(t)]$ along the random straight sections of the classical
paths in the cascade model. We like to mention that this is a highly
unreliable procedure due to the strong cancelations of terms that are
randomly generated. Our experience shows that a reliable result
(better than $<5 \%$) in the transition region $\omega\approx \Gamma$
requires an ensemble of $10^3$ cascade runs where each path has about
$10^3$ collisions.  Compared to that the analytical result
(\ref{Mtau}) has significant computational advantages, provided the
random process is of this form.

\yoursection{Quantum many-body description}
In the full quantum field theory formulation the production rate is
given by all photon self energy diagrams of skeleton type according to
the closed time-path rules \cite{LP, KB} with {\em full}
Green's function which result from the summation of
Dyson's equation. The later in short matrix notation becomes
\begin{eqnarray}\label{Dyson}
  && {\boldsymbol G}={\boldsymbol G}_0+ {\boldsymbol
   G}_0\odot{\boldsymbol\Sigma}\odot {\boldsymbol G}
   \\&&\unitlength10mm\begin{picture}(8,1.25)
   \put(0,.5){\Dysonf}\nonumber
\end{picture}\end{eqnarray}
Here ${\boldsymbol G}_0$, ${\boldsymbol G}$ and ${\boldsymbol \Sigma}$
denote the two by two matrices of the unperturbed Green's function
(thin lines), the full Green's function (full lines), and the proper
self energy of the source particles. The $\odot$ denotes the
space-time folding.  The four components of $-\ii{\boldsymbol \Sigma}$
are defined as the sum of all standard proper self energy diagrams
like in normal perturbation theory, now however with definite $+$ or
$-$ assignments at the external vertices, and summed over the $-$ and
$+$ signs for all internal vertices. These signs specify the Green's
functions linking the vertices and the choice between the $-$ vertex
$-\ii V$ and its adjoint value $+\ii V^{\dagger}$ at $+$ vertices
(c.f. ref. \cite{LP}).

The full Green's functions account for the finite damping width of the
particles, and therefore destroy of the strict energy-momentum
relation in dense matter. This width expressed through the imaginary
part $-\Gamma$ of the retarded self energy arises from collisions or
decays of the particles. In this picture the off-diagonal Green's
functions $G^{- +}$ and $G^{+ -}$ are {\em four momentum and space
Wigner densities} for the occupied and available 'single particle
states', which now have a finite width.  The corresponding
off-diagonal parts in the self energy determine the corresponding gain
and loss terms in a transport description. Therefore each diagram of
(\ref{Pi-+xq}) defines a specific partial rate.

We suggest a decomposition of the diagrams that allow for a simple
interpretation and classification in terms of physical in-medium
scattering processes, and propose particular resummations of
physically meaningful diagrams, which consider the finite damping
width of all source particles in matter.  All the graphs consisting
$G^{- +}G^{+ -}$--products are explicitly presented.  In this picture
the set of diagrams reduces to
\begin{equation}\label{keydiagrams}
   \unitlength6mm\keydiagrams\vphantom{\int^A_B}
\end{equation}
Here full dots and boxes denote effective in-medium vertices and
4-point interactions, e.g.
\begin{equation}\label{4point}
   \unitlength10mm\begin{picture}(14,.8)
   \put(0,-.6){\fourpointeq}
   \end{picture}\vphantom{\int_A^A}\\ \nonumber
\end{equation}
These in-medium vertices and four point interactions are defined for a
specific choice of sign (say $-$) through skeleton resummation
schemes, where only bare $-$ vertices linked by $G^{--}$ Green's
functions appear. Compared to conventional diagrams, vertex
corrections of different signs appear on both sides of a loop once
they are separated by $\{+-\}$ lines. In (\ref{4point}) the wavy lines
are either interactions or full boson-propagators in a theory of
fermions interacting with bosons, like in QCD. In some simplified
representations (being often used, c.f. \cite{V1}) the 4-point
functions behave like intermediate bosons (e.g. phonons).

We note that each diagram in (\ref{keydiagrams}) represents already a
whole class of perturbative diagrams of any order in the interaction
and in the number of loops. The most essential term is the one-loop
diagram in (\ref{keydiagrams}), which is positive definite, and
corresponds to the first term of the classical Langevin result for
$\Pi_{\mathrm{cl}}$ in (\ref{Mtau}).  The other diagrams represent
interference terms due to rescattering. All diagrams calculated with
full Green's functions are void of infra-red divergences. Thus these
diagrams represent the quantum generalization of the infra-red regular
Langevin result (\ref{Mtau}). Under appropriate conditions, the
correct {\em quasi--particle} (QP) and {\em quasi--classical} (QC)
limits are recovered from this subset of graphs.\\

\noindent{\em Decomposition of closed diagrams into Feynman amplitudes in the
quasi--particle approximation (QPA)}

The QPA is a quite commonly used concept originally derived for Fermi
liquids at low temperature (Landau - Migdal, see \cite{LL}), where it
constitutes a consistent scheme. With the application of transport
models to high energy situations this concept has been taken over to a
regime where its validity cannot be justified under all
circumstances. For the validity of the QPA one normally assumes that
$\Gamma\ll\bar{\epsilon}$, where $\bar{\epsilon}$ is an average
particle kinetic energy ($\sim T$ for equilibrium matter). This has
considerable computational advantages as Wigner densities ("-- +" and
"+ --" lines) become energy $\delta$--functions, and the particle
occupations can be considered to depend on momentum only rather than on
the energy variable.\footnote{The later approximation is also often
used beyond the scope of the QPA and is then known as Kadanoff--Baym
ansatz \cite{KB}, c.f.\cite{SL}}. Formally the energy integrals can be
eliminated, in diagrammatic terms just cutting the corresponding "--
+" and "+ --" lines \cite{V1}.

Thus the QP picture allows a transparent interpretation of closed
diagrams.  With consecutive numbers 1 to 6 for the diagrams drawn in
(\ref{keydiagrams}), diagrams 1, 2, 4 and 5 relate to the
radiation from a single in-medium scattering between two fermionic
quasi-particles \cite{V2}. Thereby diagram 2 is more important than
diagram 4 for neutral interactions, while this behavior reverses for
charge exchange interactions (the latter is also important for gluon
radiation in QCD transport due to color exchange interactions).
Diagrams like 3 describe the interference terms due to further
rescatterings of the source fermion with others. For diagram 6 the
photon is produced from intermediate states (its contribution is
suppressed in the soft limit). Some of the diagrams, which are not
presented explicitly in eq.(\ref{keydiagrams}) give more than two
pieces, if cut, so they do not reduce to the Feynman amplitudes. Since
one works with zero-width fermion Green's functions in QPA, the finite
width contributions have to appear in higher order diagrams through
corresponding $\Im\Sigma$-insertions! Therefore the whole set of
diagrams defining the full $-\ii\Pi^{-+}$ in QPA is by far larger,
than ours (\ref{keydiagrams}).

Moreover, our considerations show that the validity condition
$\Gamma\ll\bar{\epsilon}$ is not sufficient for the
QPA. Rather, since finally energy differences of order $\omega$ appear,
one has to demand that also $\omega\gg \Gamma$
in the QPA. In particular, the remaining series of QP-diagrams is {\em
no longer convergent} unless $\omega>\Gamma$, since arbitrary powers in
$\Gamma/\omega$ appear, and there is no hope to ever recover a reliable
result by a finite number of QP-diagrams for the production of soft
quanta! With {\em full Green's functions}, however, one
obtains a description that uniformly covers both, the soft
($\omega\ll\Gamma$) and the hard ($\Gamma\ll\omega$) regime.\\

\noindent{\em Quasi-classical limit (QCL)}

The QCL requires that $i)$ the particle occupations are small
($\left<n_{\vec p}\right> \ll 1$) implying a Boltzmann gas and that
$ii)$ all inverse length or time-scales times $\hbar$ are small
compared to the typical momentum and energy scales of the source
systems.  In particular we shall assume $\hbar \omega \ll
\bar{\epsilon}$, and a fermion collision rate $ \Gamma=\hbar/
\tau_{\mathrm{coll}} \ll\bar{\epsilon}$. The last inequality allows
to use Kadanoff--Baym ansatz, where the fermion occupations depend
only on momentum $n_{\vec p}= n_{\epsilon_{\vec p}-\mu_F}$ but no
longer on the energy $\epsilon$ of the particles.  Assuming again $
\Gamma \simeq const.$ on relevant time scales one immediately finds
\begin{equation}\label{tp}
   G^{- +}(\tau,{\vec p}) \simeq \ii n_{\vec p} \exp [-|\Gamma\tau|/2
   +\ii\epsilon_{\vec p} \tau],
\end{equation}
\begin{equation}\label{tp2}
  G^{+ -}(\tau,{\vec p}) \simeq -\ii (1-n_{\vec p})\exp [-|\Gamma\tau|/2
  -\ii\epsilon_{\vec p} \tau].
\end{equation}
With these Green functions we calculate the diagrams
(\ref{keydiagrams}). For the one--loop diagram we obtain the
expression identical to the $n=0$ term of the classical Langevin
result (\ref{Mtaun0}):
\begin{eqnarray}\label{oneloopomega}
-\ii\Pi^{-+}_0(\omega,{\vec q}) \approx  4\pi e^2 \int\frac{\di^3p}
{(2\pi)^3}
\frac{n_{{\vec p}}(1-n_{{\vec p}
})2\Gamma v^iv^k}{\left(\omega-{\vec
q}{\vec v})
\right)^2+\Gamma^2}
\approx 4\pi e^2\rho_0\left<\frac{2\Gamma v^iv^k}{\left(\omega-{\vec
q}{\vec v})
\right)^2+\Gamma^2}\right>\; .
\end{eqnarray}
For neutral interactions (corresponding to the classical examples) we
show that precisely the diagrams in the first line of
(\ref{keydiagrams})
denoted by $-\ii\Pi^{-+}_n$ with $n$ $\{-+\}$ loop insertions  correspond
to the $n$-th term of the Langevin result (\ref{Mtau})\footnote{The proof is
easily given in the $\tau-{\vec p}$ representation, assigning times $0$
and $\tau$ to the external $-$ and $+$ vertices, while the internal
$-$ and $+$ vertices are taken at $t^-_1$ to $t^-_n$ and $t^+_1$ to
$t^+_n$, respectively. In the classical limit $G^{--}$ is retarded,
while $G^{++}$ is advanced, such that both time sequences have the
same ordering: $0<t^-_1<\dots< t^-_n<\tau$ and $0<t^+_1<\dots<
t^+_n<\tau$ (the order reverses, if the line sense of the
outer fermion lines are reversed). Thus the $\tau$-dependence of the
modulus of these diagrams again results in $e^{-|\Gamma\tau|}$.
Assuming classical momentum transfers $|{\vec p}_n-{\vec p}_{n+1}|$
which are large compared to $\hbar\Gamma$ one concludes that the
$\{+-\}$ loop insertions merge $\delta (t^-_n-t^+_n)$ in their time
structure. With $t_n=t^-_n=t^+_n$ a given diagram then no longer
depends on the intermediate times $t_n$ apart from the ordering
condition, and therefore results in a factor $|\Gamma\tau|^n/n!$. With
${\vec q}=0$ also the corresponding momenta are pair-wise identical,
and the remaining momentum integrations just serve to define the
correlation between ${\vec v}_m$ and ${\vec v}_{m+n}$ after $n$
collisions of the charged particle.}. Thus
\begin{eqnarray}
-\ii\Pi^{-+}_n (\tau ,{\vec q}=0) &\approx&4\pi e^2 \rho_0\left<v^i_m
v^k_{m+n}\right>_m \frac{|\Gamma\tau|^n}{n!} e^{-|\Gamma\tau|}
\, , \\
-\ii\Pi^{-+}_n (\omega, {\vec q}=0) &\approx&4\pi e^2 \rho_0 \left<v^i_m
v^k_{m+n}\right>_m 2\Gamma^n \Re\frac{1}{(\Gamma-\ii\omega)^{n+1}}.
\end{eqnarray}

For neutral interactions all other diagrams in (\ref{keydiagrams}) drop
in the QC limit. Diagram 4 is unimportant for neutral
interactions. Other diagrams in the original series
(\ref{keydiagrams}) acquire extra powers either in the mean occupation
$\left<n_{\vec p}\right>\ll 1$ or in $\Gamma/\bar{\epsilon}\ll 1$. Such
suppression factors result from extra $G^{-+}$ lines compared to the
classical diagrams of the same topology or are due to a violation of
time ordering (diagram 5), since classical interactions are of
time-scale $1/\bar{\epsilon}$, which is short compared to the damping
time $1/\Gamma_x$.

We note that truncating the set of diagrams (\ref{keydiagrams}) with
respect to a certain expansion parameter special care should be taken in
order to satisfy charge--current conservation law.  For verification
one may use the Ward--Takahashi identities within the desired order.

For equilibrium $T\neq 0$ case even in general quantum consideration
the one--loop diagram can be expressed by the QP prescription
\cite{V1, V2} however multiplied by the factor
$C={\omega^2}/{(\omega^2+\Gamma^2)}$, which displays the suppression
at low $\omega$. There is hope that even in the quantum case some
higher order diagrams can also be resummed and that an overall
suppression factor of the form $C={\omega^2}/{(\omega^2+\Gamma_x^2)}$
emerges for the true in-matter rate relative to the quasi-free or QP
one in the limit $q=0$, c.f. the diffusion result
(\ref{Picltau}-\ref{Dqomega}).

Altogether our results provide an extension of the {\em incoherent
quasi--free}  and QP approximation from the ultra violet limit towards the
soft limit with appropriately modified production cross sections. It
does not only regularize the infra--red divergence of the free rate,
but it also produces the right $q$ dependence in the soft limit. We
also note that our diagrammatic description may suggest a formulation
of a transport theory which includes the propagation of particles with
finite width and therefore may permit a consistent treatment of
resonances in non--equilibrium dense matter.
\newpage
\noindent {\bf Acknowledgments.}

The authors acknowledge helpful discussions with G. Bertsch,
P. Daniele\-wicz, B. Friman, P. Henning and M. Herrmann.
D.N.V. thanks GSI for hospitality and support. Also the research
described in this publication was made possible for him in part by
Grant N3W000 from International Science Foundation and he thanks ISF
for Grant.\\[-10mm]


\begin{thebibliography}{99}
\itemsep=0mm
\bibitem{LPM}
L. D. Landau and I. Pomeranchuk, Dokl. Akad. Nauk SSSR {\bf 92}
(1953) 553, 735; also in Coll. Papers of Landau, ed. Ter Haar
(Gordon \& Breach, 1965) papers 75 - 77;\\
A. B. Migdal, Phys. Rev. {\bf 103}, (1956)1811;
Sov. Phys. JETP {\bf 5} (1957) 527
\bibitem{K}
J. Knoll and C. Guet, { Nucl. Phys.} {\bf A494} (1989) 334;\\
M. Durand and J. Knoll, { Nucl. Phys.} {\bf A496} (1989) 539;
\bibitem{KL}
J. Knoll and R. Lenk,
{ Nucl. Phys.} {\bf A 561} (1993) 501
\bibitem{WG}
J. Cleymans, V.V. Goloviznin, and K. Redlich,
{Phys. Rev.}{\bf D47} (1993) 989;\\
X. N. Wang and M. Gyulasssy, {Nucl. Phys.} {\bf B 420} (1994) 583;\\
R. Baier, Yu. L. Dokshitzer, S. Reigne and D. Schiff, Bielefeld BI-TP
94/57;\\
P. A. Henning and R. Sollacher, preprint, GSI-95-04
\bibitem{Proc}
Proceedings of the Tenth International Conference on Ultra--Relativistic
Nucleus--Nucleus Collisions, eds: E. Stenlund, H. --A. Gustafsson,
A.Oskarsson and I.Otterlund (North--Holland, 1994); {Nucl.Phys.} {\bf A566},
1c (1994)
\bibitem{NS}
Neutron Stars, ed. by D. Pines, R. Tamagaki and S. Tsuruta,
Addison--Weseley, N. Y. (1992)\\
G. Raffelt and D. Seckel, MPI-Ph/93-90
\bibitem{TW}
N. S. Tsamis and R. P. Woodard, {Ann. Phys.} (N. Y.) {\bf 238}
(1995) 1
\bibitem{SL}
V. Spicka and P. Lipavsky, {Phys. Rev. Lett.} {\bf 73} (1994),
3439;\\
P. Lipavsky,V. Spicka and B. Velicky, {Phys. Rev.} {\bf B34} (1986), 6933
\bibitem{D}
P. Danielewicz, {Ann. Phys.} (N. Y.) {\bf 152}
(1984) 239, 305;\\ P. Danielewicz and G. Bertsch, {Nucl. Phys.} {\bf
A533} (1991) 712
\bibitem{LP}
c.f. E. M. Lifshiz  and L. P. Pitaevskii, Physical
Kinetics (Nauka, 1979); Pergamon press (1981)
\bibitem{V1}
D. N. Voskresensky and A. V. Senatorov, {Yad. Fiz.} {\bf 45} (1987)
657;
in Engl. translation Sov.J.Nucl.Phys.{\bf 45} (1987) 411
\bibitem{Cleymans}
J. Cleymans, V.V. Goloviznin, and K. Redlich,
{Phys. Rev.}{\bf D47} (1993) 173
\bibitem{KB}
L. P. Kadanoff and G. Baym, Quantum Statistical Mechanics (Benjamin, 1962)
\bibitem{LL}
E. M. Lifshiz  and L. P. Pitaevskii, Statistical Physics, p.2
(Nauka, 1978); Pergamon press (1980);\\
A. B. Migdal, Theory of Finite Fermi Systems and Properties of
Atomic Nuclei , Willey and sons, 1967, Nauka; 1983
\bibitem{V2}
D. N. Voskresensky and A. V. Senatorov, {Yad. Fiz.} {\bf 52}
(1990) 447; in Engl. translation Sov.J.Nucl.Phys.{\bf 52 } (1990) 284
\end{thebibliography}
\end{document}